\documentstyle[preprint,aps,eqsecnum]{revtex}
\begin{document}
\draft
\def\pmb#1{\setbox0=\hbox{#1}%
     \kern-.025em\copy0\kern-\wd0
      \kern.05em\copy0\kern-\wd0
       \kern-.025em\raise.0433em\box0}
\def\btau{\pmb{$\tau$}}
\def\bsigma{\pmb{$\sigma$}}
\def\bdiamond{\pmb{$\diamond$}}
\def\bbdiamond{\pmb\bdiamond}
\def\hat{\widehat}
\def\omeg{\bmatdieci\char '41}
\title
{Relativistic Approach to \\
Superfluidity in Nuclear Matter}

\author{F. Matera, G. Fabbri and A. Dellafiore} 

\address
{Dipartimento di Fisica, Universit\`a degli Studi di Firenze\\
and Istituto Nazionale di Fisica Nucleare, Sezione di
Firenze,\\
L.go E. Fermi 2, I-50125, Firenze, Italy}

\maketitle
\begin{abstract}
~Pairing correlations in symmetric nuclear matter are studied 
within a relativistic mean--field approximation based on 
a field theory of nucleons coupled to neutral 
(~$\sigma$ and $\omega$~) and to charged (~$\varrho$~) 
mesons. The Hartree--Fock and the pairing fields are calculated 
in a self--consistent way. The energy gap is the result 
of a strong cancellation between the scalar and vector 
components of the pairing field. We find that the pair amplitude 
vanishes beyond a certain value of momentum of the paired 
nucleons. This fact determines an effective cutoff in the gap 
equation. The value of this  cutoff gives an energy 
gap in agreement with the estimates of non relativistic 
calculations. 

\end{abstract}

\pacs{PACS number(s): 21.65.+f, 21.30.Fe, 21.60.Jz}

\section{Introduction}

In the investigation of superfluidity or superconductivity phenomena 
in systems of strongly interacting fermions a crucial role is played by 
the gap equation of the BCS theory \cite{fe71}. For infinite 
homogeneous systems the kernel of the gap equation is usually 
a slowly decreasing function of the quasiparticle momentum. 
Even if the sum over momenta converges, contributions 
from terms involving large momentum components of the interaction and 
quasiparticle energies very far from the Fermi surface, are not negligible 
\cite{ba90}, despite the fact that the gap energy may result 
to be a small fraction of the Fermi energy. Contrary to what 
happens for electrons in metals, for  nuclear or neutron matter, 
there is no natural cutoff over momentum. Consequently  for 
nuclear matter a relativistic treatment is desireable. 
In the present work we investigate pairing correlations in cold 
symmetric nuclear matter by taking into account relativistic 
effects in a consistent way. A first study in this 
direction, for $^{1}S_0$ pairing in nuclear matter, has been already 
performed by H. Kucharek and P. Ring \cite{ku91} within 
the framework of the Quantum Hadrodynamics  theory 
of Ref.\cite{se86} (~QHD~). These authors, though introducing some  
simplifying assumptions (~e.g exchange terms of the interaction 
have been neglected~), have shown that pairing correlations 
and partcle--hole correlations can be described on the same foot
within  a relativistic treatment. In our opinion the consequences of 
a relativistic quantum approach have not been thoroughly exploited 
in Ref.\cite{ku91}. For instance, the amplitudes of the nucleon field 
in the quasiparticle states have been assumed to be proportional 
simply to the four--spinors of the Hartree approximation.

In a recent paper \cite{gui96} a more complete relativistic treatment 
of the Hartree--Fock--Bogoliubov approximation to QHD has been 
presented. The authors of Ref.~\cite{gui96} have shown that the pairing 
field has large scalar and vector components. The actual value of 
the gap in the excitation spectrum is the result of a strong mutual 
cancellation between these components of the pairing field. In the 
present work we obtain similar results, in particular the expression 
for the quasiparticle energy practically coincides with that 
of Ref.~\cite{gui96}. However, though the starting point of our treatment 
is the same as in Ref.~\cite{gui96}, i.e. a relativistic generalization of 
the Gorkov scheme \cite{gor58}, we do not need to introduce 
any particular "ansatz" for an effective single--particle 
Lagrangian, as done in Ref.~\cite{gui96}. Moreover, we show that 
a more consistent treatment, from a relativistic point 
of view, brings out a new interesting feature: the pair 
amplitude vanishes beyond a certain momentum of the paired 
nucleons. The value of this momentum depends on the nuclear 
density and on the strength of the self--consistent 
pairing field. As a consequence in the relativistic gap equation 
a natural cutoff over momentum occurs. In the previous works of 
Refs.~\cite{ku91,gui96} it has been necessary to introduce such a 
cutoff in an arbitrary way, in order to obtain a satisfactory 
agreement with the current estimates of nonrelativistic 
calculations \cite{ba90,che86,ku89,ain89,ba92,ba94,elg96} 
for the energy gap.

We base  our approach on the version I of QHD (~QHD--I~) 
\cite{se86}, we include also the charged $\varrho$ meson field 
in a phenomenological way similar to the ~MFT~ approach to ~QHD--II~ 
of Ref.~\cite{se86}. For simplicity we neglect the 
pion field. Adding the pions would produce only 
small quantitative differences in our results \cite{gui96}. 

Here we are mainly interested in studying new effects introduced 
by a relativistic approach to superfluidity. First we perform  
our calculations by using the same approximations already 
introduced in Refs.~\cite{del91} and \cite{mat94} for studying 
collective modes and response functions of nuclear matter. 
These approximations amount to neglecting finite--range and 
retardation effects in the exchange of mesons 
between nucleons. Within  this approach we obtain simple expressions 
for the relevant quantities, where we can easily appreciate the role 
played by the various ingredients of the theory. In spite of 
the approximations introduced, the essential features of a 
relativistically covariant treatment are retained.

The value of the energy gap obtained with the approach just 
mentioned is much larger than the value predicted by 
nonrelativistic treatments \cite{ba90,che86,ku89,ain89,ba92,ba94,elg96}. 
A better quantitative agreement can be achieved by taking 
into account the finite range of the nucleon--nucleon interaction. 
Retardation effects instead do not play a significant role. We 
evaluate the finite--range effects by using an iterative procedure 
and show that the first iteration gives already a sufficiently accurate 
approximation.\hfill\break

\section{Formalism}

In the model adopted here the nucleons are coupled to neutral mesons 
(~$\sigma$ and $\omega$~) and to the charged vector meson, 
$\varrho$. According to the usual procedure employed in studying 
superconductivity we consider ensembles with indefinite number of particles. 
Therefore we add to the Lagrangian the term 
$\mu~\bar{\psi}(x){\gamma^0}\psi(x)~$, where $\mu$ is the chemical 
potential  (~$\psi(x)$ denotes the 8--component nucleon field~). 
In the end the value of the chemical potential will be 
determined by fixing the average baryon density. With these 
ingredients the field equations are \cite{se86}
\begin{mathletters}
\label{all2.1}
\begin{equation}
\bigl(i{\partial_\lambda}\gamma^\lambda-{g_V}V_\lambda(x)\gamma^\lambda-
g_\varrho{\bf B}_\lambda(x)\cdot\btau\gamma^\lambda+\mu\gamma^0-M+{g_S}
\Phi(x)\bigr)\psi(x)=0~,\label{2.1a}
\end{equation}
\begin{equation}
\bigl({\partial_\lambda}{\partial^\lambda}+{m_S}^2\bigr)\Phi(x)=
{g_S}\bar{\psi}(x)\psi(x)~,\label{2.1b}
\end{equation}
\begin{equation}
{\partial_\lambda}W^{\lambda\nu}(x)+{m_V}^2{V^\nu}(x)=
{g_V}\bar{\psi}(x){\gamma^\nu}\psi(x)~,\label{2.1c}
\end{equation}
\begin{equation}
{\partial_\lambda}{\bf L}^{\lambda\nu}(x)+{m_\varrho}^2{{\bf B}^\nu}(x)=
{g_\varrho}\bar{\psi}(x){\gamma^\nu}\btau\psi(x)~,\label{2.1d}
\end{equation}
\end{mathletters}
where $W^{\lambda\nu}(x)={\partial^\lambda}{V^\nu}(x)-
{\partial^\nu}{V^\lambda}(x)~$ 
and ${\bf L}^{\lambda\nu}(x)={\partial^\lambda}
{{\bf B}^\nu}(x)-{\partial^\nu}{{\bf B}^\lambda}(x)~$. 
The quantities $\Phi(x)$, ${V^\nu}(x)$ and ${{\bf B}^\nu}(x)$ represent 
the scalar, vector and charged vector fields, respectively. In the 
present paper we are concerned with a uniform system at equilibrium, 
it is convenient to study such a system through the one--particle 
density matrix in four--momentum space:
\begin{equation}
{\hat F}_{\alpha\beta}(p)=
{1\over(2\pi)^4}\int d^4R~e^{-ip\cdot R}
<0|:\bar{\psi}_\beta({R\over2})\psi_\alpha(-{R\over2}):|0>~,\label{2.2}
\end{equation}
where $\alpha$ and $\beta$ are double indices for spin and isospin. 
The dots denote normal ordering and $|0>$ is the correlated ground state.
Since we consider only symmetric nuclear matter the 
matrix ${\hat F}(p)$ is diagonal and degenerate with respect to 
isospin indices. 

By using the procedures outlined in Ref.~\cite{del91},
the following equation for ${\hat F}(p)$ can be derived from 
Eqs.~(\ref{all2.1}) 
\begin{eqnarray}
&[p&_\lambda{\gamma^\lambda}+\mu\gamma^0{\hat F}(p)]_{\alpha\beta}-
M{\hat F}_{\alpha\beta}(p)
\nonumber \\
&-&g_V\frac{1}{(2\pi)^4}\int d^4R~e^{-ip\cdot R}
<0|:\bar{\psi}_\beta(\frac{R}{2}){\gamma^\lambda_{\alpha\delta}}\psi_\delta
(-\frac{R}{2})V_\lambda(-\frac{R}{2}):|0>
\nonumber \\
&-&g_{\varrho}\frac{1}{(2\pi)^4}\int d^4R~e^{-ip\cdot R}
<0|:\bar{\psi}_\beta(\frac{R}{2}){[\gamma^\lambda\btau]_{\alpha\delta}}
\psi_\delta(-\frac{R}{2})\cdot
{\bf B}_\lambda(-\frac{R}{2}):|0>
\nonumber \\
&+&g_S\frac{1}{(2\pi)^4}\int d^4R~e^{-ip\cdot R}
<0|:\bar{\psi}_\beta(\frac{R}{2})\psi_\alpha(-\frac{R}{2})
\Phi(-\frac{R}{2}):|0>=0~.\label{2.3}
\end{eqnarray}

Following Refs.~\cite{del91} and \cite{mat94} we neglect derivative 
terms in Eqs.~(\ref{2.1b},{c},{d}) so that the meson field operators 
are simply given by 
\begin{eqnarray}
\Phi(x)&=&\frac{g_S}{m_S^2}\bar\psi(x)\psi(x)~,
\nonumber \\
V^\lambda(x)&=&\frac{g_V}{m_V^2}\bar\psi(x){\gamma^\lambda}\psi(x)~,
\nonumber \\
{\bf B}^\lambda(x)&=&\frac{g_\varrho}{m^2_\varrho}\,\bar{\psi}(x)
\gamma^\lambda\btau\psi(x)
~.\label{2.4}
\end{eqnarray}
This approximation simplifies calculations considerably, 
however it neglects retardation and finite--range effects 
in the exchange of mesons between nucleons. Nevertheless, because 
of the small Compton wavelengths of the heavy mesons 
with respect to the internucleon spacing in ordinary nuclear matter, 
the approximation (\ref{2.4}) appears to be reasonable. 
Clearly, for pions this approximation would not be justified.

After substituting in Eq.~(\ref{2.3}) the expressions (\ref{2.4}) for the 
meson field operators we obtain an equation which contains 
expectation values of products of four nucleon field operators: 
\begin{equation}
<0|:\bar\psi_\beta(\frac{R}{2})\psi_\alpha(-\frac{R}{2})
\bar\psi_\gamma(-\frac{R}{2})\psi_\delta(-\frac{R}{2}):|0>~.
\label{2.5}
\end{equation}
Following Gorkov \cite{gor58} these quantities are approximated 
by two--fold products of expectation values: 
\begin{eqnarray}
&<0&|:\bar\psi_\beta(\frac{R}{2})\psi_\alpha(-\frac{R}{2})
\bar\psi_\gamma(-\frac{R}{2})\psi_\delta(-\frac{R}{2}):|0>=
\nonumber \\
&<0&|:\bar\psi_\beta(\frac{R}{2})
\psi_\alpha(-\frac{R}{2}):|0><0|:\bar\psi_\gamma(-\frac{R}{2})
\psi_\delta(-\frac{R}{2}):|0>
\nonumber \\
&-&<0|:\bar\psi_\beta(\frac{R}{2})\psi_\delta(-\frac{R}{2}):|0>
<0|:\bar\psi_\gamma(-\frac{R}{2})\psi_\alpha(-\frac{R}{2}):|0>
\nonumber \\
&+&<0|:\bar\psi_\beta(\frac{R}{2})\bar\psi_\gamma(-\frac{R}{2}):|0>
<0|:\psi_\delta(-\frac{R}{2})\psi_\alpha(-\frac{R}{2}):|0>~.
\label{2.6}
\end{eqnarray}
The first two terms correspond to the Hartree--Fock field. Beside these 
terms, Eq.~(\ref{2.6}) contains the product of elements of the 
anomalous density matrix. A nonvanishing value of this product 
implies the presence of the superfluid phase.

Introducing the pair amplitude 
\begin{equation}
{\hat D}_{\alpha\beta}(p)=
{1\over(2\pi)^4}\int d^4R~e^{-ip\cdot R}
<0|:\bar{\psi}_\beta(\frac{R}{2})\bar\psi_\alpha(-\frac{R}{2}):|0>~,
\label{2.7}
\end{equation}
\noindent
and the pairing field
\begin{equation}
{\hat\Delta}_{\alpha\beta}=
\int d^4p~{\hat D}_{\alpha\beta}(p)=
<0|:\bar\psi_\beta(x)\bar\psi_\alpha(x):|0>~,\label{2.8}
\end{equation}
we obtain the following equation for the one--particle density matrix
\begin{eqnarray}
&\bigl(&p_\lambda\gamma^\lambda-{\tilde f}_V{\rho_B}\gamma^0+
\mu\gamma^0\bigr)
{\hat F}(p)-\bigl(M-{\tilde f}_S\rho_S\bigr){\hat F}(p)
\nonumber \\
&-&f_V\gamma^\lambda{\hat{\tilde \Delta}}\gamma_{\lambda}^T{\hat D}(p)
-f_\varrho\gamma^\lambda\btau{\hat{\tilde \Delta}}
\cdot\btau^T\gamma_{\lambda}^T{\hat D}(p)
+f_S{\hat{\tilde \Delta}}{\hat D}(p)
=0~,\label{2.9}
\end{eqnarray}
where the effective coupling constants, $\tilde f_S$ and $\tilde f_V$, are 
given by the combinations
\begin{equation}
{\tilde f}_S=\frac{7}{8}f_S+\frac{1}{2}f_V+\frac{3}{2}f_{\varrho}~,~~~~~~~ 
{\tilde f}_V=\frac{1}{8}f_S+\frac{5}{4}f_V+\frac{3}{4}f_{\varrho}
~,\label{2.10}
\end{equation}
with $f_S=(~{g_S}/{m_S}~)^2$,  $f_V=(~{g_V}/{m_V}~)^2$ and 
$f_{\varrho}=(~{g_{\varrho}}/{m_{\varrho}}~)^2$. 
\noindent
The matrix $\hat{\tilde \Delta}$ in (\ref{2.9})is conjugate to 
the pairing field:
\begin{equation}
{\hat{\tilde \Delta}}_{\alpha\beta}=~\bigl(\gamma^0 {\hat \Delta}^\dagger
\gamma^0\bigr)_{\alpha\beta}=~<0|:\psi_\beta(x)\psi_\alpha(x):|0>
~.\label{2.11}
\end{equation}
The scalar density $\rho_S$ and the baryon density $\rho_B$ are given by 
\begin{eqnarray}
&\rho_S&=2Tr\int d^4p\,{\hat F}(p)~,
\nonumber \\
&\rho_B&=2Tr\int d^4p\,\gamma^0{\hat F}(p)~,\label{2.12}
\end{eqnarray}
with the factor $2$ coming from isospin degeneracy; the traces 
are taken only over the spin states. 

By repeating the same procedure that leads to Eq.~(\ref{2.9}), 
we obtain for the pair amplitude ${\hat D}(p)$ an equation 
coupled to Eq.~(\ref{2.9}), that reads 
\begin{eqnarray}
&\bigl(&p_\lambda\gamma^\lambda+{\tilde f}_V{\rho_B}\gamma^0-
\mu\gamma^0\bigr)
{\hat D}(p)-\bigl(M-{\tilde f}_S\rho_S\bigr){\hat D}(p)
\nonumber \\
&-&f_V{\gamma^\lambda}{\hat{ \Delta}}\gamma_{\lambda}^T{\hat F}(p)
-f_\varrho{\gamma^\lambda}\btau{\hat \Delta}
\cdot\btau^T\gamma_{\lambda}^T{\hat F}(p)
+f_S{\hat \Delta}{\hat F}(p)
=0~,\label{2.13}
\end{eqnarray}

In Eqs.~(\ref{2.9}) and (\ref{2.13})  the exchange contributions 
to the mean field have been taken into account through the 
definition of  the effective coupling constants of Eq.~(\ref{2.10}). 

Before turning our attention to Eqs.~(\ref{2.9}) and (\ref{2.13}), 
we discuss the symmetry and tensor properties of the pairing 
field. Here we consider only ${^1}S_0$ pairing of nucleons 
in symmetric nuclear matter. In this case  the pairing field 
$\hat \Delta$ is symmetric and degenerate with 
respect the isospin indices. Therefore we can 
consider only the isoscalar component of $\hat \Delta$ and hence omit the 
isospin variables. Now $\hat \Delta$ is a $4\times 4$ matrix in 
spin space like the matrices ${\hat F}(p)$ and ${\hat D}(p)$. 
In spin space the matrix $\hat \Delta$ is antisymmetric, 
hence it can be decomposed as 
\begin{equation}
{\hat \Delta}=~{\Delta_S}\sigma^{13}+{\Delta_{PS}}\sigma^{02}
+{\Delta_0}\gamma^5\gamma^2+{\Delta_1}\gamma^3+
{\Delta_2}\gamma^5\gamma^0
+{\Delta_3}\gamma^1~,\label{2.14}
\end{equation}
the subscripts are chosen according to the tensor properties of the 
various terms. The transformations properties 
of $\hat \Delta$ can be derived from the basic transformation 
law of the field operator $\psi (x)$. In detail, under 
infinitesimal Lorentz transformations (~$\Lambda^{\lambda\nu}=
g^{\lambda\nu}+\epsilon^{\lambda\nu}$~) the field $\hat \Delta$ transforms 
according to 
\[
{\hat \Delta}^\prime=~{\hat \Delta}+{i\over 4}\epsilon^{\lambda\nu}
({\hat \Delta}\sigma_{\lambda\nu}+
{\sigma_{\lambda\nu}}^T{\hat \Delta})~,
\]
while for space inversion 
\[
{\hat \Delta}^\prime=~\gamma^0{\hat \Delta}\gamma^0~.
\]
From these equations we can see that the four quantities $({\hat \Delta}_0, 
-i{\hat \Delta}_1, {\hat \Delta}_2, i{\hat \Delta}_3)$ 
represent a four--vector 
and the remaining components ${\hat \Delta}_S$ and ${\hat \Delta}_{PS}$
are scalar and pseudoscalar quantities, respectively. Since we 
are considering a homogeneous system at rest, only 
the components ${\Delta}_S$ and ${\Delta}_0$ in Eq.~(\ref{2.14}) 
can differ from zero: 
\begin{equation}
{\hat \Delta}=~{\Delta_S}\sigma^{13}
+{\Delta_0}\gamma^5\gamma^2~.\label{2.15}
\end{equation}
Moreover, because of the invariance of the equilibrium 
state under time--reversal the components ${\Delta_S}$ 
and ${\Delta_0}$ can be assumed to be real and 
the matrix ${\hat{\tilde \Delta}}$ of Eq.~(\ref{2.11}) 
becomes 
\begin{equation}
{\hat{\tilde \Delta}}=~{\hat \Delta}~.\label{2.16}
\end{equation}

From Eqs.~(\ref{2.9}) and (\ref{2.13}), after some algebraic 
manipulations, we obtain two separate equations 
for the density matrix ${\hat F}(p)$ and 
for the pair amplitude ${\hat D}(p)$:
\begin{equation}
\bigl(R^\mu(p)\gamma_\mu-\overline M(p)\bigr)\,\hat F(p)=0~\label{2.17}
\end{equation}
and 
\begin{equation}
\bigl(R^\mu(-p)\gamma_\mu^T-\overline M(-p)\bigr)\,\hat D(p)=0~.
\label{2.18}
\end{equation}
The components of the four--vector $R_\mu(p)$ are 
\begin{eqnarray}
&{\bf R}&=\,{\bf p}~,
\nonumber \\
&R&_0(p)=\,p_0+{\tilde \mu}+{2\over{{\cal K}(p)}}\,\bigl(g^2\Delta^2_S
\tilde \mu-f^2\Delta^2_0 p_0+fg\Delta_S\Delta_0M^*\bigr)~,
\label{2.19}
\end{eqnarray}
and the mass term $\overline M(p)$ is given by
\begin{equation}
\overline M(p)\,=\,M^*-{2\over{{\cal K}(p)}}\,\bigl(f^2\Delta^2_0\,M^*-
f\,g\Delta_0\,\Delta_S(p_0-{\tilde \mu})\bigr)~,\label{2.20}
\end{equation}
Here the coupling constants $g$ and $f$ are given by the combinations
\begin{equation}
g=\,f_S-4\,(f_V+f-\varrho)~,~~~~~~f=\,f_S+2\,(f_V+f_\varrho)~,
\label{2.21}
\end{equation}
and the quantity ${\cal K}(p)$ is expressed as 
\begin{equation}
{\cal K}(p)=\,\bigl(p_0-\tilde\mu\bigr)^2-E^2_{\bf p}+f^2\,\Delta^2_0-
g^2\,\Delta^2_S~,\label{2.22}
\end{equation}
where $E_{\bf p}=({\bf p}^2+M^{*2})^{1/2}$, while  
${\tilde\mu}=\mu-{\tilde f}_V\rho_B$ is the effective chemical 
potential and $M^*=M-{\tilde f}_S\rho_S$ is the effective nucleon mass. 

Equation (\ref{2.17}) tells us that the matrix ${\hat F}(p)$ can be 
put in the form
\begin{mathletters}
\label{all2.23}
\begin{equation}
{\hat F}(p)=\,F(p)+\gamma ^\mu F_\mu(p)~,
\label{2.23a}
\end{equation}
with
\begin{equation}
F_\mu(p)=\,{{R_\mu(p)}\over{\overline M(p)}}F(p)~,\label{2.23b}
\end{equation}
\end{mathletters}
and $F(p)$ a scalar quantity. 

Equations (\ref{2.17}) and (\ref{2.18}) contain the components 
of the pairing field as parameters, these components must be 
determined self--consistently. This can be done by 
solving Eq.~(\ref{2.13}) with respect to ${\hat D}(p)$ with the aid 
of Eqs.~(\ref{all2.23}). For the two nonvanishing components 
of the pair amplitude 
\[
{\hat D}(p)=\,D_S(p)\sigma^{13}+D_0(p)\gamma^5\gamma^2
\]
we obtain 
\begin{mathletters}
\label{all2.24}
\begin{equation}
D_S(p)=\,{2\over{{\cal K}(p)}}\,\bigl(g\Delta_S\,\tilde\mu F_0(p)-
f\Delta_0(p_0F(p)-M^*F_0(p))\bigr)~,\label{2.24a}
\end{equation}
\begin{equation}
D_0(p)=\,{2\over{{\cal K}(p)}}\,\bigl(g\Delta_S\,\tilde\mu F(p)-
f\Delta_0(p_0F_0(p)-M^*F(p))\bigr)~.\label{2.24b}
\end{equation}
\end{mathletters}

Now we derive the energy spectrum using our approach. 
Substituting in  Eq.~(\ref{2.17}) the formal solution 
given by Eqs.~(\ref{all2.23}), we can see that the 
components of the four--vector $R_\mu(p)$ must satisfy 
the constraint 
\[
R_\mu R^\mu=\,{\overline M}^2(p)~.
\]
This equation, for a fixed $\bf p$, determine the allowed values 
of $p_0$
\begin{eqnarray}
p_0^2=\,&E&_{\bf p}^2+{\tilde\mu}^2+g^2\Delta_S^2
+f^2\Delta_0^2~\pm 
\nonumber \\
&2&\,\bigl({\tilde\mu}^2E_{\bf p}^2+g^2f^2\Delta_S^2\Delta_0^2+f^2\Delta_0^2
{\bf p}^2-2fg\,\Delta_S\Delta_0\,{\tilde\mu}M^*\bigr)^{1/2}~,
\label{2.25}
\end{eqnarray}
which correspond to the energies of elementary excitations 
referred to the effective chemical potential $\tilde\mu$. 
The upper sign refers to elementary excitations of the 
Dirac sea. Actually in the absence of the pairing field the 
energies of these excitations become  
\[
p_0=\,(E_{\bf p}+{\tilde\mu})~.
\]
We neglect contributions from the Dirac sea and 
consider only the energy value given by the 
lower sign in Eq.~(\ref{2.25}). This defines the energy 
${\cal E}_{\bf p}$ of a quasiparticle with momentum 
${\bf p}$ in the superfluid phase. 

We turn now to the explicit evaluation of the matrices 
${\hat F}(p)$ and ${\hat D}(p)$. By inserting a complete 
set of energy eigenvectors in Eq.~(\ref{2.2}) we have 
\begin{equation}
{\hat F}_{\alpha\beta}(p)=
{1\over(2\pi)^3}\int d^3{\bf R}~e^{i{\bf p\cdot R}}
\sum_n\delta(p_0+E_n)<0|\bar{\psi}_\beta({{\bf R}\over2})|n>
<n|\psi_\alpha(-{{\bf R}\over2})|0>~,\label{2.26}
\end{equation}
where $|n>$ represents a quasiparticle state 
$|n>\equiv |{\cal E}_{{\bf p}_n},{\bf p}_n,\lambda_n>$ 
with spin label $\lambda_n$. This equation, toghether with 
Eq.~(\ref{2.17}), shows that the one--particle density matrix 
${\hat F}(p)$ can be put in the form 
\begin{equation}
{\hat F}_{\alpha\beta}(p)=
{1\over(2\pi)^3}\sum_\lambda\,g^*_\lambda({\bf p})
{\bar u}^{(\beta)}_\lambda({\bf p})u^{(\alpha)}_\lambda({\bf p}) 
g_\lambda({\bf p})\delta\bigl(p_0+{\cal E}_{\bf p}\bigr)~,\label{2.27}
\end{equation}
where the spinor  $u_\lambda({\bf p})$ obeys the equation 
\begin{equation}
\bigl(R_\mu({\bf p},-{\cal E}_{\bf p})\gamma^\mu-
{\overline M}(-{\cal E}_{\bf p})\bigr)u_\lambda({\bf p})
=\,0~.\label{2.28}
\end{equation}
With the normalization $u^*_\lambda({\bf p})u_\lambda({\bf p})=1$, 
the quantity $g_\lambda({\bf p})$ represents  the 
probability amplitude of  finding a hole with momentum $-{\bf p}$ 
and spin label $-\lambda$ in the quasiparticle state 
$|{\cal E}_{\bf p},{\bf p},\lambda>$. 

For the pair amplitude ${\hat D}(p)$, with the same procedure and 
with the aid of Eq.~(\ref{2.18}), we obtain the expression 
\begin{equation}
{\hat D}_{\alpha\beta}(p)=
{1\over(2\pi)^3}\sum_\lambda\,(-1)^\lambda g^*_{-\lambda}({\bf p})
{\bar u}^{(\beta)}_\lambda({\bf p}){\bar v}^{(\alpha)}_\lambda({\bf p}) 
{\tilde g}_\lambda({\bf p})\delta\bigl(p_0+{\cal E}_{\bf p}\bigr)~.
\label{2.29}
\end{equation}
The equation for the spinor $v_\lambda({\bf p})$ can be derived from 
Eq.~(\ref{2.18}) and reads 
\begin{equation}
\bigl(R_\mu(-{\bf p},{\cal E}_{\bf p})\gamma^\mu-
{\overline M}({\cal E}_{\bf p})\bigr)v_\lambda({\bf p})
=\,0~.\label{2.30}
\end{equation}
We choose for $v_\lambda({\bf p})$ the same normalization  
as $u_\lambda({\bf p})$. In Eq.~(\ref{2.30}) 
the quantity ${\tilde g}_\lambda({\bf p})$ is the probability amplitude of  
finding a particle with momentum ${\bf p}$ and spin label $\lambda$ 
in the quasiparticle state $|{\cal E}_{\bf p},{\bf p},\lambda>$.  
With the choice of phase made in Eq.~(\ref{2.29}) the product 
$g^*_{-\lambda}({\bf p}){\tilde g}_\lambda({\bf p})$ is 
real and independent of $\lambda$, as can be seen by  
substituting in Eq.~(\ref{2.13}) the expression (\ref{2.27}) 
for ${\hat F}(p)$.

It is implicit in our approach that a quasiparticle state is a 
superposition of one--particle and one--hole states, so that 
the following normalization condition holds 
\begin{equation}
\vert g_\lambda({\bf p})\vert^2+\vert {\tilde g}_\lambda({\bf p})\vert^2
=\,1~.\label{2.31}
\end{equation}
Moreover, since we are considering a homogeneous and isotropic system, 
$\vert g_\lambda({\bf p})\vert$ and 
$\vert{\tilde g}_\lambda({\bf p})\vert$ do not depend on $\lambda$. 

The specific form of the spinors $u_\lambda({\bf p})$ and 
$v_\lambda({\bf p})$ is determined by the signs of $R_0(p)$ 
and ${\overline M}(p)$. Explicit calculations show that 
$R_0(\pm {\cal E}_{\bf p})$ remains positive 
for any value of $\vert {\bf p}\vert$, whereas the mass term 
${\overline M}(-{\cal E}_{\bf p})$ of Eq.~(\ref{2.28}) has 
a peculiar behaviour. It is a monotonically decreasing 
function of $\vert {\bf p}\vert$, it takes negative values for 
$\vert {\bf p}\vert$ larger than a certain value $p_c$ and becomes 
infinitely negative at a finite value of $\vert {\bf p}\vert>p_c$. 
The mass term ${\overline M}({\cal E}_{\bf p})$ 
of Eq.~(\ref{2.30}) instead is always positive  and almost constant. 

For $\vert{\bf p}\vert\leq p_c$ the solutions of Eq.~(\ref{2.28}) 
are given by
\begin{mathletters}
\label{all2.32} 
\begin{equation}
u_\lambda({\bf p})=\,\Bigl[{{R_0(-{\cal E}_{\bf p})+
{\overline M}(-{\cal E}_{\bf p})}\over{2R_0(-{\cal E}_{\bf p})}}\Bigr]^{1/2}
\left(\matrix{\chi_\lambda \cr
~~\cr
\displaystyle{ 
{{\bsigma\cdot{\bf p}}\over{R_0(-{\cal E}_{\bf p})+
{\overline M}(-{\cal E}_{\bf p})}}}\,\chi_\lambda\cr}\right)~,\label{2.32a}
\end{equation}
whereas for $\vert{\bf p}\vert>p_c$ we have to choose 
the spinors
\begin{equation} 
u_\lambda({\bf p})=\,\Bigl[{{R_0(-{\cal E}_{\bf p})+\vert
{\overline M}(-{\cal E}_{\bf p})\vert}\over{2R_0(-{\cal E}_{\bf p})}}
\Bigr]^{1/2}
\left(\matrix{\displaystyle{ 
{{\bsigma\cdot{\bf p}}\over{R_0(-{\cal E}_{\bf p})+
\vert{\overline M}(-{\cal E}_{\bf p})\vert}}}\,\chi_\lambda \cr
~~\cr
\chi_\lambda\cr}\right)~,\label{2.32b}
\end{equation}
\end{mathletters}
as solutions of Eq.~(\ref{2.28}). The solutions of Eq.~(\ref{2.30}) 
are given by the spinors 
\begin{equation}
v_\lambda({\bf p})=\,\Bigl[{{R_0({\cal E}_{\bf p})+
{\overline M}({\cal E}_{\bf p})}\over{2R_0({\cal E}_{\bf p})}}\Bigr]^{1/2}
\left(\matrix{\chi_\lambda \cr
~~\cr
\displaystyle{- 
{{\bsigma\cdot{\bf p}}\over{R_0({\cal E}_{\bf p})+
{\overline M}({\cal E}_{\bf p})}}}\,\chi_\lambda\cr}\right)~.\label{2.33}
\end{equation}
In equations above $\chi_\lambda$ denote the usual two--component 
Pauli spinors. 

We remark that the energy spectrum of excitations of the Dirac sea 
remains well separated from the quasiparticle spectrum. In fact the
mass terms in Eqs.~(\ref{2.17}) and (\ref{2.18}) for the antiparticle case,  
are always positive for any value of $\vert {\bf p}\vert$. 

By substituting the spinors (\ref{2.32b}) and (\ref{2.33}) 
in Eq.~(\ref{2.29}), we can see that for $\vert{\bf p}\vert> p_c$, 
the two components 
\[
D_S({\bf p})=\,{1\over 4}Tr\sigma^{13}{\hat D}({\bf p})
\]
and
\[
D_0({\bf p})=\,{1\over 4}Tr\gamma^5\gamma^2{\hat D}({\bf p})
\]
vanish. This implies that $F_0({\bf p})$ also vanishes for 
$\vert{\bf p}\vert> p_c$, see Eqs.~(\ref{all2.24}). 
The occupation number of particles in the correlated ground state 
displays a discontinuity. 

From Eqs~(\ref{all2.24}) with the aid of the normalization 
condition (\ref{2.31}), we can determine the amplitudes 
$g_\lambda({\bf p})$ and ${\tilde g}_\lambda({\bf p})$, 
and the matrices ${\hat F}(p)$ and ${\hat D}(p)$, 
as functions of the parameters $M^*$, $\tilde\mu$, 
$\Delta_S$ and $\Delta_0$. These parameters can be calculated 
by solving the four coupled equations that are obtained 
by fixing the baryon density 
\begin{mathletters}
\label{all2.34}
\begin{equation}
{2\over 3}\,{1\over{\pi^2}}p_F^3=\,8\int d{\bf p}F_0({\bf p})~,
\label{2.34a}
\end{equation}
and using the self--consistency relations for the effective 
nucleon mass 
\begin{equation}
M^*=\,M-8{\tilde f}_S\int d{\bf p}{{{\overline M}(-{\cal E}_{\bf p})}
\over{R_0(-{\cal E}_{\bf p})}}\,F_0({\bf p})\label{2.34b}
\end{equation}
and for the components of the pairing field 
\begin{equation}
\Delta_S=\,2\int {{d{\bf p}}\over{{\cal K}({\bf p},
-{\cal E}_{\bf p})}}\,\bigl(g\Delta_S\,\tilde\mu F_0({\bf p})-
f\Delta_0(-{\cal E}_{\bf p}F({\bf p})-M^*F_0({\bf p}))\bigr)~,
\label{2.34c}
\end{equation}
\begin{equation}
\Delta_0=\,2\int {{d{\bf p}}\over{{\cal K}({\bf p},
-{\cal E}_{\bf p})}}\,\bigl(g\Delta_S\,\tilde\mu F({\bf p})-
f\Delta_0(-{\cal E}_{\bf p}F_0({\bf p})-M^*F({\bf p}))\bigr)~.
\label{2.34d}
\end{equation}
\end{mathletters}

The momentum $p_c$, for which $F_0({\bf p})$ vanishes, 
plays the role of an effective cutoff in the integrals 
(\ref{all2.34}). This fact avoids introducing an 
artificial cutoff to make the integrals (\ref{2.34c}) and 
(\ref{2.34d}) converge.  The dependence of $p_c$ on $\Delta_S$ 
and $\Delta_0$ amounts to a further self--consistency relation. 

Equations (\ref{2.34c}) and (\ref{2.34d}) replace the gap 
equation of the nonrelativistic case. In the nonrelativistic 
limit $F({\bf p})\rightarrow F_0({\bf p})$, hence these two 
equations become identical and the two components of the pairing  
field coincide: $\Delta_S=\Delta_0$. Moreover, the quasiparticle 
energy ${\cal E}_{\bf p}$ acquires the usual expression 
of the BCS theory
\[
{\cal E}_{\bf p}=\,\bigl((E_{\bf p}-{\tilde\mu})^2+(f+g)^2\Delta^2
\bigr)^{1/2}~,
\]
where $\Delta=\Delta_S=\Delta_0$. 

\section{ Results}

In this section we investigate the solutions of Eqs.~(\ref{all2.34}). 
These equations have been derived by starting from the assumptions 
(\ref{2.4}), which neglect the finite range of the nucleon--
nucleon interaction. 

Concerning the coupling constants $f_V$ and $f_S$ we choose their value 
so as to reproduce the binding energy (~$15.75~MeV$~) of saturated nuclear 
matter with a Fermi momentum of $1.42~~fm^{-1}$ 
(~see Ref.\cite{del91}~). These values are:
\[
f_S=2.37{\cdot 10^{-4}}MeV^{-2},~~
f_V=1.45{\cdot 10^{-4}}MeV^{-2}~.
\]
For the coupling constant $f_{\varrho}$ we have taken the value determined 
by the $\varrho\rightarrow2\pi$ decay 
\[
f_\varrho=1.55{\cdot 10^{-5}}MeV^{-2}~.
\]
Then the effective constants $f$ and $g$ for the pairing field 
are 
\[
f=5.58{\cdot 10^{-4}}MeV^{-2},~~
g=-4.05{\cdot 10^{-4}}MeV^{-2}~.
\]

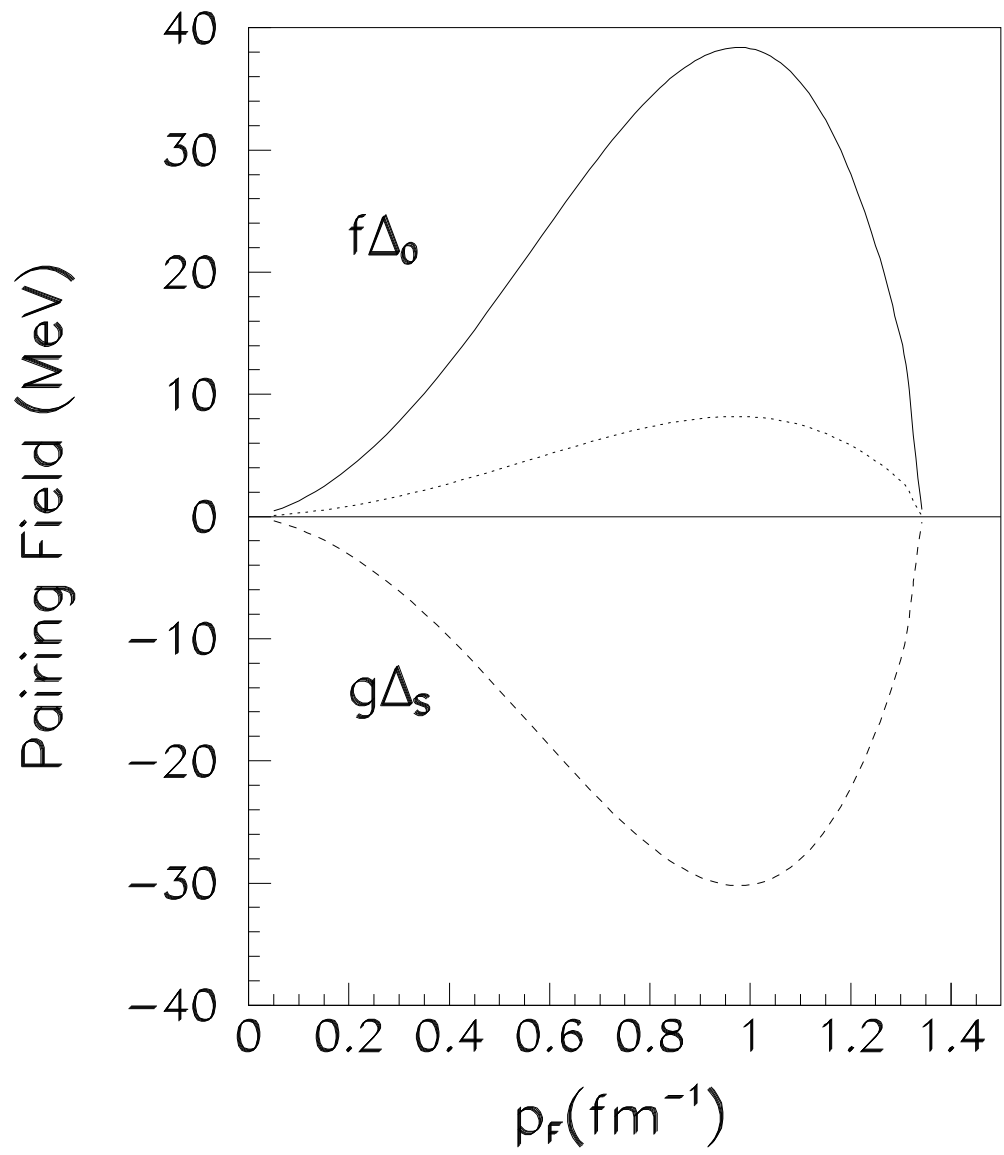
\begin{figure}
\caption{Components of the pairing field (~times 
the respective coupling constants~) as a function of $p_F$. The solid 
line and the dashed line correspond to the vector and the scalar 
components, respectively. The dotted line gives the sum of 
these two quantities.}
\label{f:1}
\end{figure}

The relevant quantities for the quasiparticle energy spectrum are 
$f\,\Delta_0$ and $g\,\Delta_S$. In Fig.~\ref{f:1} 
these quantities are displayed, together with their sum, as 
functions of the Fermi momentum. The sum $f\,\Delta_0+g\,\Delta_S$ 
approximatively reproduces the gap in the quasiparticle 
energy spectrum. Figure~\ref{f:1} shows that a superfluid solution 
of Eqs.~(\ref{all2.34}) is present in the range of 
densities corresponding to $p_F<1.35\,fm^{-1}$. Moreover one can 
see that $f\,\Delta_0$ and $g\,\Delta_S$ separately are very large 
with respect to their sum, i.e. the gap in the quasiparticle 
spectrum is determined by difference between two large 
and not very different numbers. 

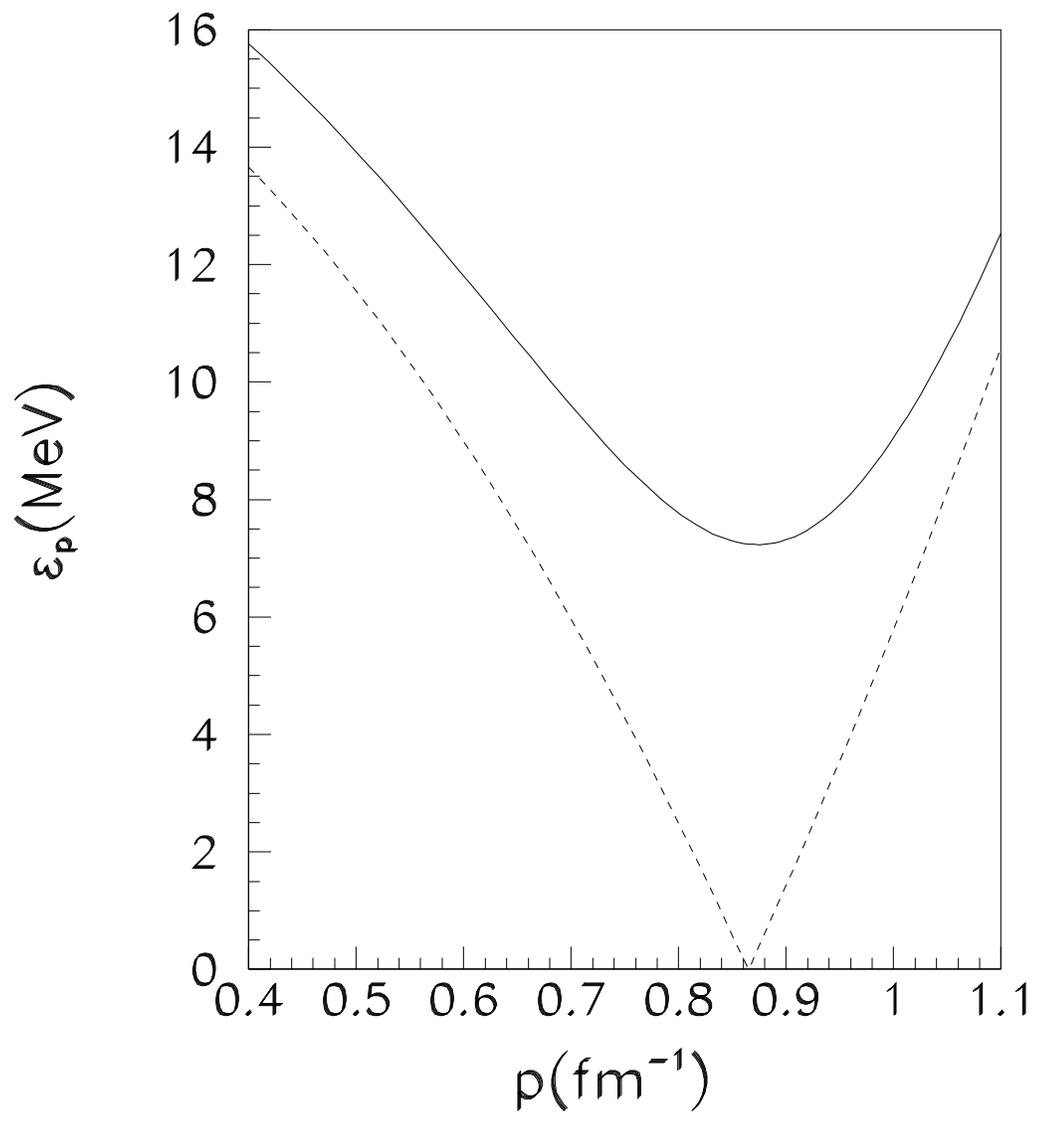
\begin{figure}
\caption{Quasiparticle energy (~solid line~) and 
single--particle energy $E_{\bf p}-\tilde\mu$ (~dashed line~), 
for $p_F=\,0.9 fm^{-1}$.}
\label{f:2}
\end{figure}

In Fig.~\ref{f:2} the excitation spectra of the superfluid and normal 
phases are displayed as functions of $\vert{\bf p}\vert$ for 
$p_F=0.9\,fm^{-1}$. A numerical analysis shows that 
for $\vert {\bf p}\vert>\sim 1.9\,fm^{-1}$ 
the quasiparticle energies lie slightly below the 
excitation energies of the normal phase, $E_{\bf p}-\tilde\mu$. 
This fact gives rise to the cutoff in the 
integrals (\ref{all2.34}). Since for a sufficiently high value of 
$\vert{\bf p}\vert$ the quantity ${\cal K}(p)$ of Eq.~(\ref{2.22}) 
may vanish, the second term of ${\overline M}(p)$, Eq.~(\ref{2.20}), 
which is positive, can become larger than $M^*$. 
The cutoff $p_c$ is given by the value of $\vert{\bf p}\vert$ 
for which the r.h.s of Eq.~(\ref{2.20}) vanishes.

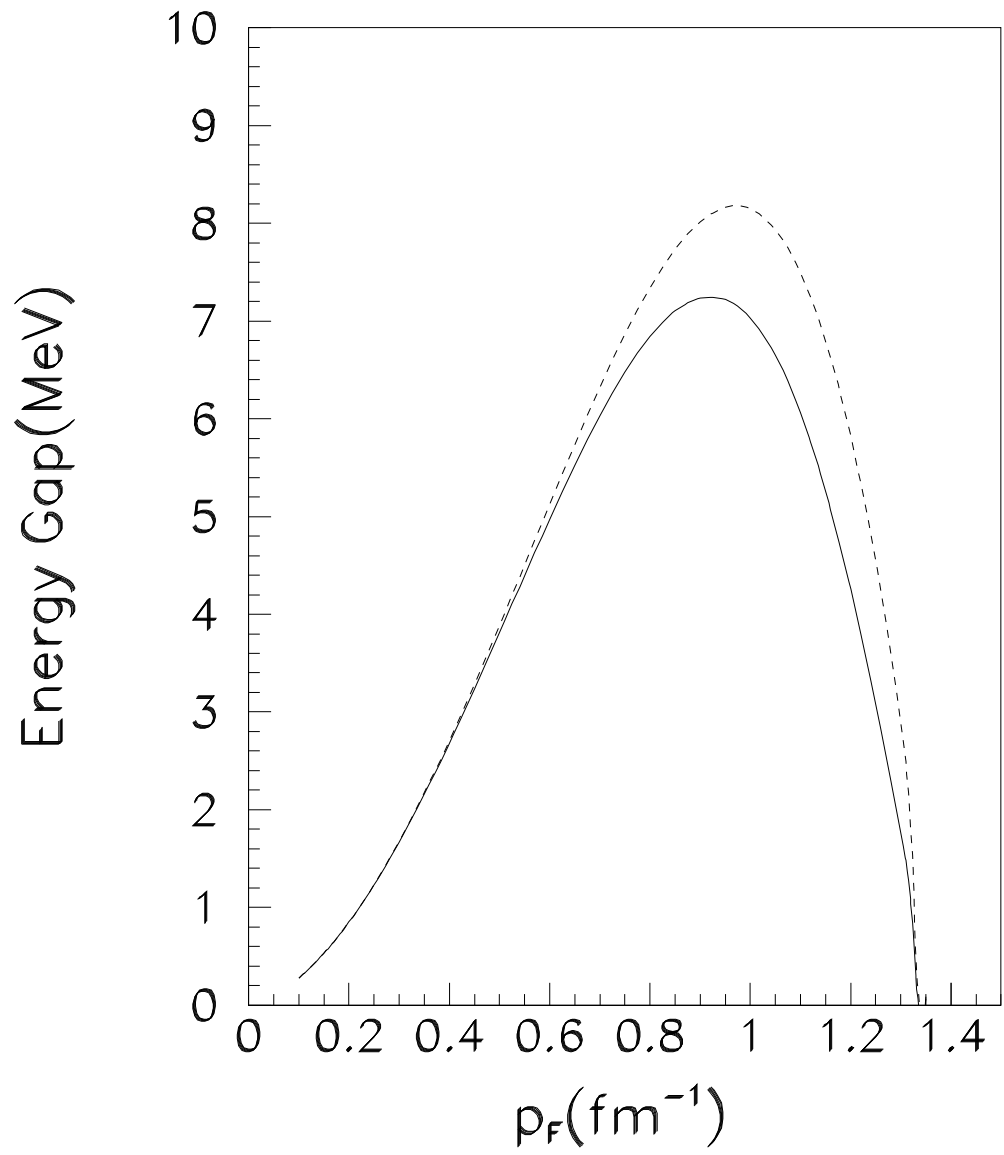
\begin{figure}
\caption{Energy gap (~solid line~) together with 
the sum $f\Delta_0+g\Delta_S$ (~dashed line~) versus 
$p_F$.}
\label{f:3}
\end{figure}

In Fig.~\ref{f:3} we show both the sum $f\,\Delta_0+g\,\Delta_S$ 
and the energy gap as a function of the Fermi momentum. 
We can see that, though remaining small, the difference 
between these two quantities increases with $p_F$. 
This is because relativistic effects become more important 
with increasing density. In the non relativistic limit 
the energy gap and the sum $f\,\Delta_0+g\,\Delta_S$ 
coincide. Then, the difference between these two quantities could 
give an insight about the relevance of relativistic effects. 
In the region around $p_F=0.9\,fm^{-1}$ where the energy gap 
takes its maximum value, this difference is not very large, 
only about $10$\%. However we remark that the occurrence of a 
cutoff in the integrals for the pairing field is obtained only 
using a relativistic expression (~Eq.~(\ref{2.25})~) for the 
quasiparticle energy.

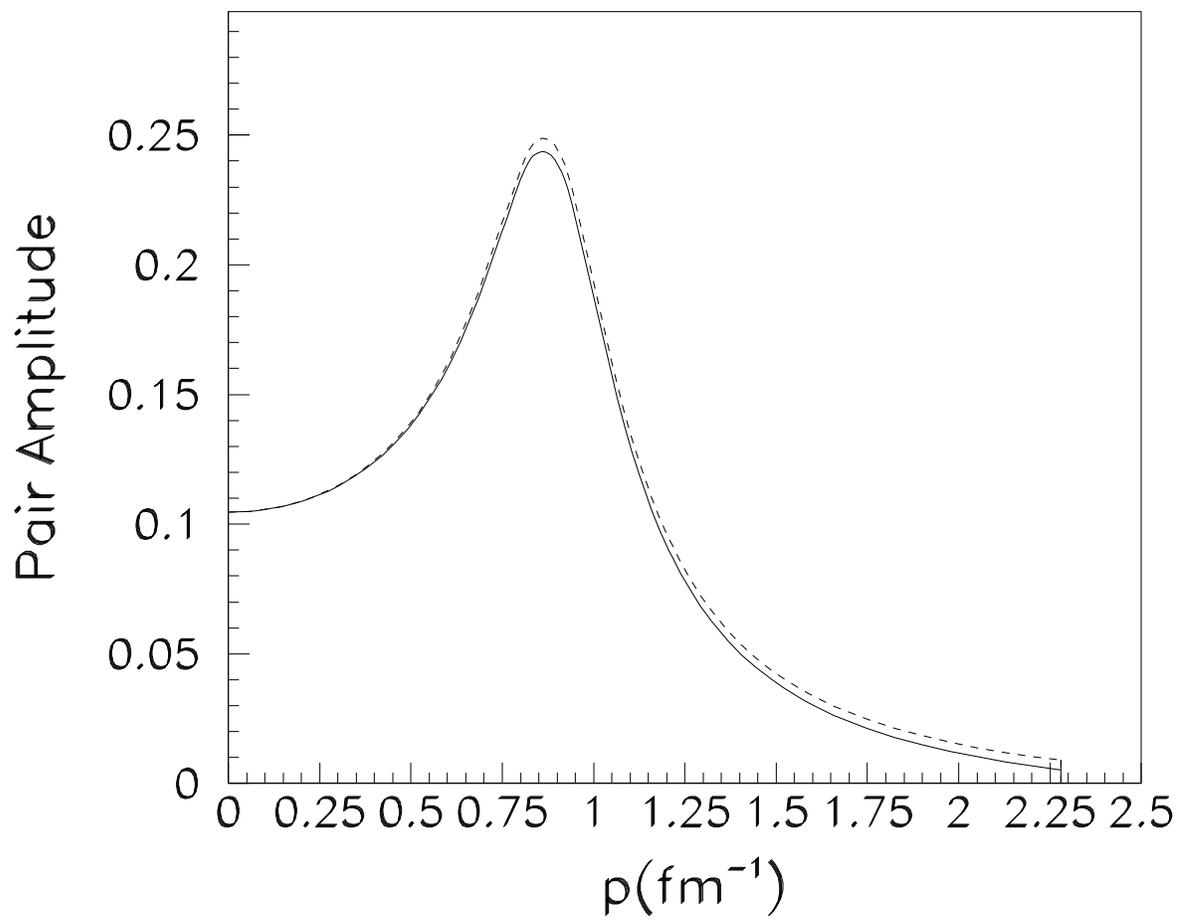
\begin{figure}
\caption{Components of the pair amplitude (~times $(2\pi\hbar)^3$~) 
as a function of $\vert{\bf p}\vert$ for $p_F=\,0.9 fm^{-1}$. 
The solid and the dashed lines correspond to the vector 
and scalar components, respectively.}
\label{f:4}
\end{figure}

Figure~\ref{f:4} shows that the pair amplitude is rather sharply peaked 
about the Fermi momentum. This important feature is the basis 
of the approximation, that we use in the next section to take 
into account the finite range of the nucleon--nucleon interaction.

For the two values of the Fermi momentum  for which superfluidity 
disappears and the energy gap becomes maximum, our 
results qualitatively agree with previous treatments, both 
nonrelativistic \cite{ba90,che86,ku89,ain89,ba92,ba94,elg96} 
and relativistic \cite{ku91,gui96}. Instead for the most relevant 
quantity of the superfluid phase, the energy gap, our 
calculations yield values that are twice the generally 
accepted estimates. The value of the sum $f+g$, 
which plays the role of an effective coupling constant for interacting 
paired nucleons is too large. It is worth noticing  
that the role of $f+g$ in determining the energy gap is enhanced 
by a cooperative effect due to the self--cosistency constraint for the 
cutoff $p_c$: if the gap becomes  larger the value of $p_c$ 
increases, then the contribution to the pairing field of the 
integrals (\ref{2.34c}) and (\ref{2.34d}) also increases, 
giving rise to a larger gap.

\section{Finite--Range Effects}
In this section we evaluate effects due to the finite 
range of the nucleon--nucleon interaction. For the effects 
of retardation in the propagation of the meson fields  
we will give only an estimate of their magnitude and 
argue that these effects do not play an important 
role. For simplicity we take into account the 
finite range of the effective interaction between paired nucleons 
only when calculating quantities that are 
relevant to the superfluid phase. In deriving the Hartree--Fock 
field, instead, we retain the approximation of Eqs.~(\ref{2.4}). 
Thus for quantities containing the pairing field, 
instead of the approximated expressions (\ref{2.4}) 
we introduce the formal solutions of Eqs.~(\ref{2.1b},{c},{d}): 
\begin{eqnarray}
\Phi(x)&=&g_S\int d^4y\,{\cal D}(x-y)\bar\psi(y)\psi(y)~,
\nonumber \\
V_\lambda(x)&=&g_V\int d^4y\,{\cal D}^{(\omega)}_{\lambda\mu}(x-y)
\bar\psi(y){\gamma^\mu}\psi(y)~,
\nonumber \\
{\bf B}_\lambda(x)&=&g_\varrho\int d^4y\,{\cal D}^{(\varrho)}_{\lambda\mu}(x-y)
\bar{\psi}(y)\gamma^\mu\btau\psi(y)~
~,\label{4.1}
\end{eqnarray}
where ${\cal D}(x-y)$, ${\cal D}^{(\omega)}_{\lambda\mu}(x-y)$ and 
${\cal D}^{(\varrho)}_{\lambda\mu}(x-y)$ are the propagators of 
the meson fields. 

We explicitly derive the equations for the contributions 
of the scalar field alone, for the vector fields only the 
final results are reported.

In Eq.~(\ref{2.6}) the term containing elements of the 
anomalous density matrix acquires the form: 
\[
\int d^4y\,{\cal D}(y-{R\over2})
<0|:\bar\psi_\beta({R\over2})\bar\psi_\gamma(y):|0>
<0|:\psi_\gamma(y)\psi_\alpha(-{R\over2}):|0>~,
\]
then in Eq.~(\ref{2.9}) the term 
$f_S{\hat{\tilde \Delta}}\,{\hat D}(p)$ is replaced by 
\begin{equation}
g_S^2\int d^4q\,{\cal D}(p-q)\gamma^0 {\hat D}^\dagger (q)\gamma^0\,
{\hat D}(p)~.\label{4.2}
\end{equation}
Here ${\cal D}(p-q)$ is the Fourier transform of the 
propagator ${\cal D}(x-y)$. 

Analogously, in Eq.~(\ref{2.13}) we have to make the substitution 
\begin{equation}
f_S{\hat \Delta}\,{\hat F}(p)\rightarrow\,g_S^2\int d^4q\,{\cal D}(p-q)
{\hat D}(q)\,{\hat F}(p)~.\label{4.3}
\end{equation}
Assuming that retardation effects are negligible, we can put 
$p_0-q_0=\,0$ in evaluating the integrals of Eqs.~(\ref{4.2}) and 
(\ref{4.3}). Thus these integrals depend only on ${\bf p}$. This 
fact greatly simplifies calculations. In particular, the 
quasiparticle energies can still be expressed in closed form; 
it is sufficient to replace in Eq.~(\ref{2.25}) the terms 
$g\Delta_S$ and $f\Delta_0$ with the analogous 
${\bf p}$--dependent quantities.

We introduce now a further and more fundamental approximation. 
In the previous section we have shown that the components of 
the pair amplitude ${\hat D}(q)$ are strongly peaked 
about $p_F$. The width of this  peak is 
much smaller than the range over which the meson propagators 
present an appreciable variation. In fact this 
range is typically of order $\sim\,m_S, m_V, m_\varrho$. 
For this reason we expect that the values of the integrals 
(\ref{4.2}) and (\ref{4.3}) can be given by 
\begin{equation}
g_S^2\,{\overline {\cal D}}(\vert{\bf p}\vert,p_F)\int d^4q\,
\gamma^0{\hat D}^\dagger(q)\gamma^0=\,
g_S^2\,{\overline {\cal D}}(\vert{\bf p}\vert,p_F)\gamma^0
{\hat \Delta}^\dagger\gamma^0~,\label{4.4}
\end{equation}
\begin{equation}
g_S^2\,{\overline {\cal D}}(\vert{\bf p}\vert,p_F)\int d^4q\,
{\hat D}(q)=\,g_S^2\,{\overline {\cal D}}(\vert{\bf p}\vert,p_F)
{\hat \Delta}~,\label{4.5}
\end{equation}
with a satisfactory approximation. Here 
${\overline {\cal D}}(\vert{\bf p}\vert,p_F)$ stands for the 
average over the directions of $\bf q$. Thus the quantity 
$f_S\hat\Delta$ in Eqs.~(\ref{2.9}) and (\ref{2.13}) 
is simply replaced by 
\begin{equation}
f_S\,a_S(\vert{\bf p}\vert,p_F)\,\hat\Delta~,\label{4.6}
\end{equation}
where the factor
\[
a_S(\vert{\bf p}\vert,p_F)=\,m_S^2
{\overline {\cal D}}(\vert{\bf p}\vert,p_F)=\,{m_S^2
\over{4p_F\vert{\bf p}\vert}}\,\ln\biggl({{m_S^2+
(\vert{\bf p}\vert+p_F)^2}\over{m_S^2+(\vert{\bf p}\vert-p_F)^2}}\biggr)
\]
represents the finite--range correction to the contribution of 
the scalar meson.

As far as the vector mesons are concerned, with some 
algebraic manipulations we can see that the matrices 
$f_V\gamma^\lambda{\hat \Delta}\gamma_\lambda^T$ and 
$f_V\gamma^\lambda\btau{\hat \Delta}\cdot\btau\gamma_\lambda^T$ 
in Eqs.~(\ref{2.9})and (\ref{2.13}) must be replaced by 
\begin{mathletters}
\label{all4.7}
\begin{equation}
4f_V\bigl({3\over4}a_V(\vert{\bf p}\vert,p_F)+{1\over4}\bigr)\Delta_S
\sigma^{13}-2f_V\bigl({1\over4}a_V(\vert{\bf p}\vert,p_F)+
{1\over2}\bigr)\Delta_0\gamma^5\gamma^0\label{4.7a}
\end{equation}
for the $\omega$ meson, and by a similar expression 
for the $\varrho$ meson, 
\begin{equation}
4f_\varrho\bigl({3\over4}a_\varrho(\vert{\bf p}\vert,p_F)+{1\over4}\bigr)
\Delta_S\sigma^{13}-2f_\varrho\bigl({1\over4}a_\varrho(\vert{\bf p}
\vert,p_F)+{1\over2}\bigr)\Delta_0\gamma^5\gamma^0~.\label{4.7b}
\end{equation}
\end{mathletters}
Here 
\[
a_V(\vert{\bf p}\vert,p_F)=\,m_V^2
{\overline {\cal D}}(\vert{\bf p}\vert,p_F)=\,{m_V^2
\over{4p_F\vert{\bf p}\vert}}\,\ln\biggl({{m_V^2+
(\vert{\bf p}\vert+p_F)^2}\over{m_V^2+(\vert{\bf p}\vert-p_F)^2}}\biggr)~,
\]
\[
a_\varrho(\vert{\bf p}\vert,p_F)=\,m_\varrho^2
{\overline {\cal D}}(\vert{\bf p}\vert,p_F)=\,{m_\varrho^2
\over{4p_F\vert{\bf p}\vert}}\,\ln\biggl({{m_\varrho^2+
(\vert{\bf p}\vert+p_F)^2}\over{m_\varrho^2+(\vert{\bf p}\vert-p_F)^2}}
\biggr)~.
\]
The correction factors are given by the bracketed terms in 
Eqs.~(\ref{all4.7}). 

Summarizing, in the present approximation the finite range 
of the nucleon--nucleon interaction is taken 
into account by replacing in all the equations of Sect. II the 
pairing coupling constants $f$ and $g$, with the combinations 
\begin{mathletters}
\label{all4.8}
\begin{equation}
a_f(\vert{\bf p}\vert,p_F)=\,f_S\,a_S(\vert{\bf p}\vert,p_F)
+f_V\bigl(a_V(\vert{\bf p}\vert,p_F)+1\bigr)+f_\varrho\bigl(
a_\varrho(\vert{\bf p}\vert,p_F)+1\bigr)\label{4.8a}
\end{equation}
and
\begin{equation}
a_g(\vert{\bf p}\vert,p_F)=\,f_S\,a_S(\vert{\bf p}\vert,p_F)
-f_V\bigl(3a_V(\vert{\bf p}\vert,p_F)+1\bigr)-f_\varrho\bigl(
3a_\varrho(\vert{\bf p}\vert,p_F)+1\bigr)\label{4.8b}
\end{equation}
\end{mathletters}
respectively.  

The pairing coupling constants, separately considered, are not 
much affected by the finite--range corrections. In fact, 
the values of $a_f(\vert{\bf p}\vert,p_F)$  and 
$a_g(\vert{\bf p}\vert,p_F)$ for $\vert{\bf p}\vert=p_f$, with 
$0.6\,fm^{-1}<p_F<1.0\,fm^{-1}$, are smaller than $f$ and $g$ 
by about $10$\% and $2$\%, respectively. However, 
for the sum $f+g$, that practically determines the magnitude 
of the energy gap, the correction is more important. 
In the same range of $p_F$ the value of 
$a_f(p_F,p_F)+a_g(p_F,p_F)$  is about ${2\over3}(f+g)$. 

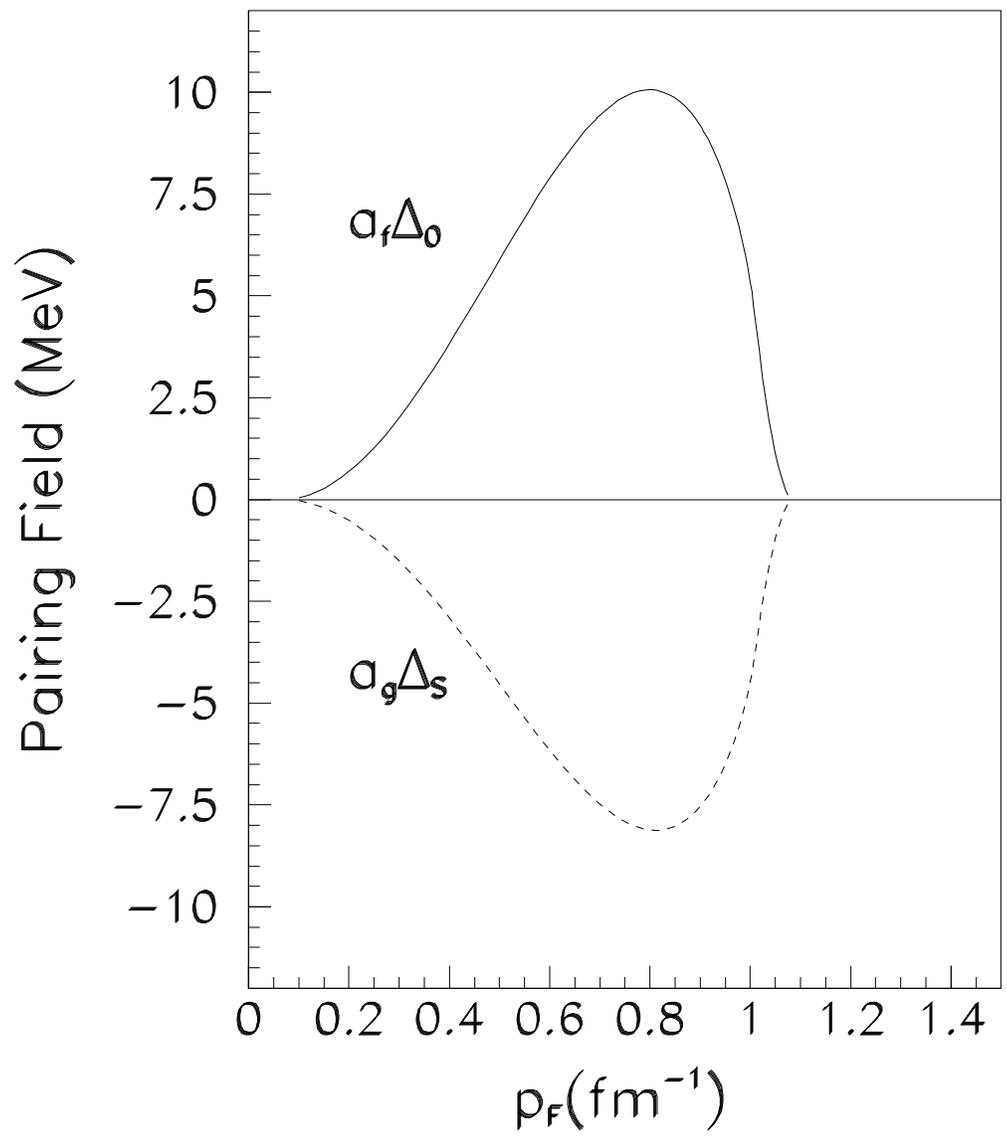
\begin{figure}
\caption{Components of the pairing field at the Fermi surface 
as functions of $p_F$. Finite--range corrections are 
included. The vector component (~solid line~) and the scalar 
component (~dashed line~) are respectively multiplied by 
$a_f$ and $a_g$ (~see Eqs.~(\ref{all4.8})~).}
\label{f:5}
\end{figure}

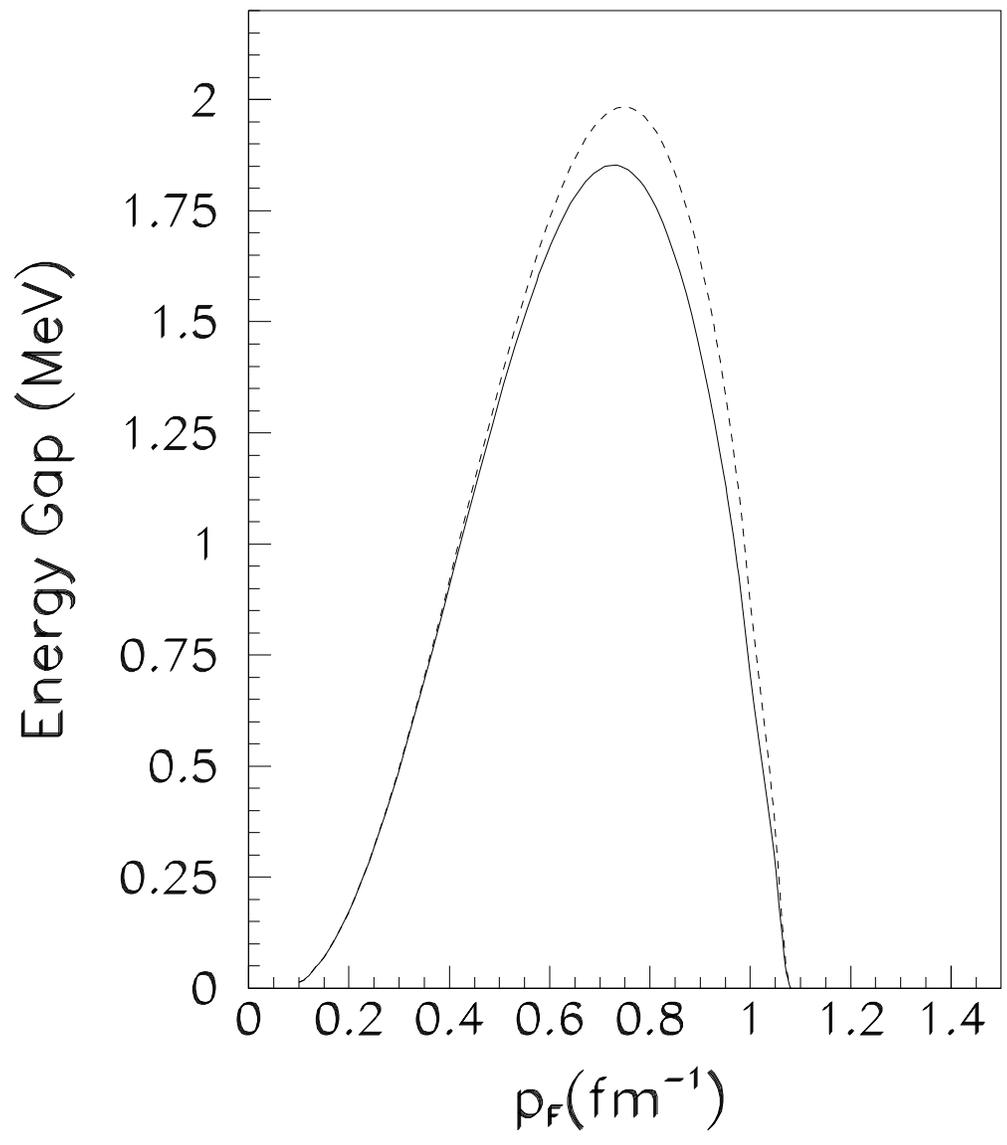
\begin{figure}
\caption{Energy gap (~solid line~) versus $p_F$ together 
with the sum $a_f\Delta_0+a_g\Delta_S$, 
evaluated at the Fermi surface (~dashed line~). The finite--range 
corrections are included.}
\label{f:6}
\end{figure}

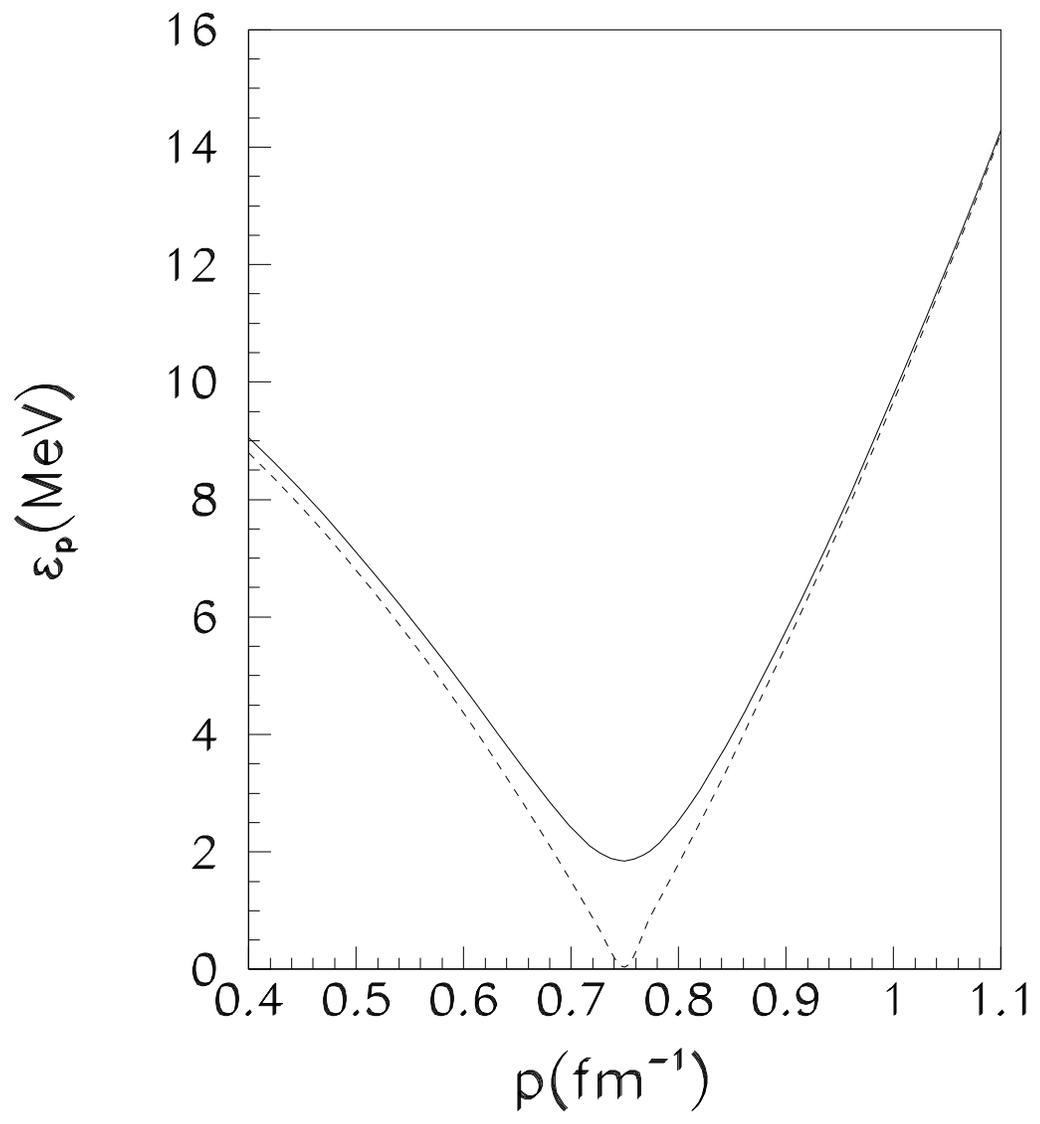
\begin{figure}
\caption{Same as Fig. 2  with finite--range 
corrections and for $p_F=\,0.8 fm^{-1}$.}
\label{f:7}
\end{figure}

The most important effects of the finite--range corrections are 
a substantial quenching of the pairing field and a reduction of 
the energy gap by an overall factor three or more. 
This is shown in Figs.~\ref{f:5} and \ref{f:6} where 
the calculated components of the pairing field and of the energy 
gap are shown as functions of $p_F$. Moreover, with respect to the 
zero--range approximation the domain of $p_F$ 
where the superfluid phase can arise, is narrowed 
and the maximum of the energy gap is shifted towards 
a lower value of $p_F$,  

In Fig.~\ref{f:7} the quasiparticle energy together with the 
excitation spectrum of the normal phase is displayed for 
$p_F=\,0.8fm^{-1}$, where the energy gap takes now its maximum value. 
The value of $\vert{\bf p}\vert$ where the two curves cross, corresponds to 
the cutoff $p_c$. A numerical analysis shows 
that $p_c\sim\,1.7fm^{-1}$.

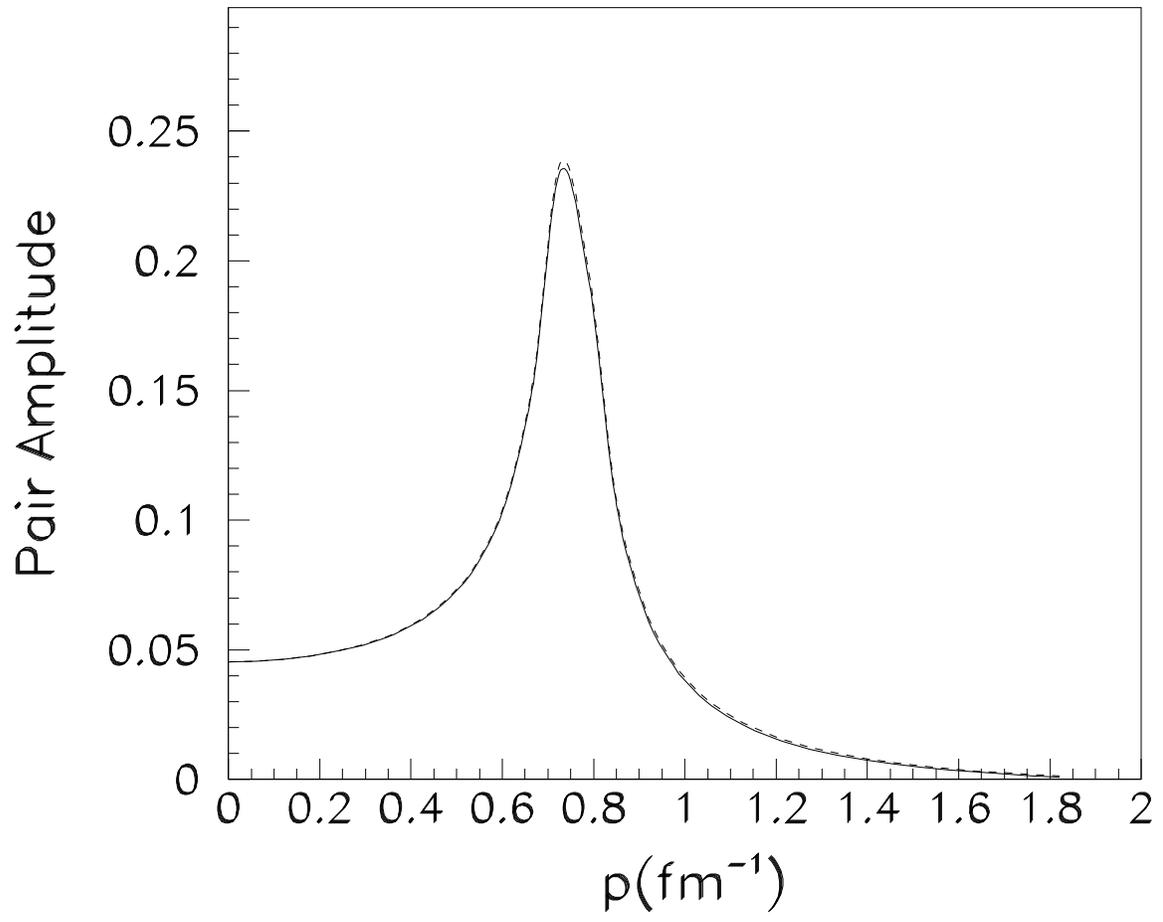
\begin{figure}
\caption{Same as Fig. 4 with finite--range 
corrections and for $p_F=\,0.8 fm^{-1}$. The two components 
of the pair amplitude are practically indistinguishable at this 
density.}
\label{f:8}
\end{figure}

Finally, in Fig.~\ref{f:8} the components of the pair amplitude are 
shown as functions of $\vert{\bf p}\vert$ for $p_F=\,0.8fm^{-1}$. 
The finite--range corrections makes the pair amplitude even  
more peaked about $p_F$. This fact further justifies the 
approximation expressed by Eqs.~(\ref{4.4}) and (\ref{4.5}). 
That approximation simplifies calculations substantially. 
In fact, the basic quantities, i.e. the components of the pairing 
field, that are determined self--consistently, are still two constants. 
We have assessed the reliability of this approximation by 
successive iterations of Eqs.~(\ref{4.2}) and (\ref{4.3}), 
starting from Eqs.~(\ref{4.4}) and (\ref{4.5}). The 
correction coming from the first iteration amounts to a few 
percent, while the second iteration does not substantially 
modify the first--order results.

We have evaluated also the order of magnitude of retardation 
effects, replacing ${\cal E}_{\bf p}$ and ${\cal E}_{\bf q}$ 
instead of $p_0$ and $q_0$ in Eqs.~(\ref{4.2}) and (\ref{4.3}). 
The change is less than one percent.

\section{Summary}
We have investigated in the framework of a relativistic model, 
the possibility for the onset of a superfluid phase 
in symmetric nuclear matter and the features of this phase. 
We have derived equations for the relevant quantities 
of the superfluid phase by making a relativistic 
generalization of the scheme introduced by Gorkov to study 
superconductivity in electron systems. In this scheme nuclear 
matter is described as an ensemble of quasiparticles 
moving in the Hartree--Fock field plus a self--consistent 
paring field. At first both the Hartree--Fock field and the 
pairing field have been treated on the same foot by using 
an approximation where the finite range and the 
retardation of the meson propagation between nucleons have 
been neglected. Then, we have improved our approach 
by introducing finite--range effects for the quantities pertaining 
to the superfluid phase. 

In the relativistic treatment of the pairing process 
for a system at rest the pairing field has two different 
components: a Lorentz scalar $\Delta_S$ and the time--component 
$\Delta_0$ of a four--vector. The different behaviour of 
the two components under Lorentz transformations must be 
properly taken into account in the relativistic 
hydrodynamics of nuclear systems in the superfluid phase.

In the non relativistic limit $\Delta_S$ and $\Delta_0$ coincide. 
In our approach they are slightly different. This does not 
mean, however, that relativistic effects are negligible. 
Actually, the expression for the quasiparticle energy derived 
in our calculations, differs from the nonrelativistic 
expression of the BCS theory in an essential way. 
The quasiparticle energy given by Eq.~(\ref{2.25}) or by its 
analogous expression that includes finite--range corrections, 
displays the salient feature that it can be less than 
$(E_{\bf p}-{\tilde \mu})$ beyond a certain value of 
$\vert{\bf p}\vert$. This determines the occurence of a cutoff in 
the relativistic gap equations (\ref{2.34c}) and (\ref{2.34d}), 
that does not appear in the analogous nonrelativistic 
expression. This cutoff removes the contributions of high 
$\vert{\bf p}\vert$ components of the interaction to 
the gap equation, thus allowing the use of nucleon--nucleon 
interactions that are only slowly decreasing for high values 
of $\vert{\bf p}\vert$. This fact reduces the energy gap appreciably.

\end{document}